\documentclass[reprint,superscriptaddress,amsmath,amssymb,aps,onecolumn,]{revtex4-2}

\usepackage[utf8]{inputenc}
\usepackage[T1]{fontenc}
\usepackage{bm}
\usepackage{subfig}
\usepackage{upgreek}

\usepackage{graphicx}% Include figure files
\usepackage{dcolumn}% Align table columns on decimal point
\usepackage{bm}% bold math
\usepackage{hyperref}% add hypertext capabilities
\usepackage{mathrsfs}
\usepackage{subfig}
\usepackage{color}
\usepackage{sansmath}
\usepackage{multirow}

\DeclareMathAlphabet{\mathsfit}{T1}{\sfdefault}{\mddefault}{\sldefault}
\SetMathAlphabet{\mathsfit}{bold}{T1}{\sfdefault}{\bfdefault}{\sldefault}

\begin{document}

\title{Active particles crossing sharp viscosity gradients}

\author{Jiahao Gong}
\affiliation{Department of Mathematics,\\
University of British Columbia, Vancouver, BC, Canada}
\author{Vaseem A. Shaik}
\affiliation{Department of Mechanical Engineering,\\
University of British Columbia, Vancouver, BC, Canada}
\author{Gwynn J. Elfring}%
 \email{gelfring@mech.ubc.ca}
\affiliation{Department of Mathematics,\\
University of British Columbia, Vancouver, BC, Canada}
\affiliation{Department of Mechanical Engineering,\\
University of British Columbia, Vancouver, BC, Canada}

\date{\today}% It is always \today, today,

\begin{abstract}
Active particles (living or synthetic) often move through inhomogeneous environments, such as gradients in light, heat or nutrient concentration, that can lead to directed motion (or \textit{taxis}). Recent research has explored inhomogeneity in the rheological properties of a suspending fluid, in particular viscosity, as a mechanical (rather than biological) mechanism for taxis. Theoretical and experimental studies have shown that gradients in viscosity can lead to reorientation due to asymmetric viscous forces. In particular, recent experiments with {\textit{Chlamydomonas reinhardtii}} algae swimming across sharp viscosity gradients have observed that the microorganisms are redirected and scattered due to the viscosity change. Here we develop a simple theoretical model to explain these experiments. We model the swimmers as spherical squirmers and focus on small, but sharp, viscosity changes. We derive a law, analogous to Snell's law of refraction, that governs the orientation of active particles in the presence of a viscosity interface. Theoretical predictions show good agreement with experiments and provide a mechanistic understanding of the observed reorientation process.
\end{abstract}

\maketitle

\section{Introduction}
Active particles are living or non-living entities that convert stored energy to directed motion and a suspension of these particles is termed active matter \cite{Schweitzer2007}. Examples of active particles range from nanorobots and microorganisms to birds, fish and even humans \cite{Toner1995, Toner1998}. Our focus here is on micron-sized active particles that move through a viscous fluid such that inertia is negligible. Active particles at this scale exhibit rich phenomena like the boundary accumulation \cite{Li2009, Berke2008}, upstream swimming \cite{Hill2007, Kaya2009, Kaya2012, Zottl2012, Peng2020}, collective motion \cite{Toner2005}, active turbulence \cite{Alert2022} and motility-induced phase separation \cite{Cates2015}.

Active particles often move through inhomogeneous environments with spatial gradients in light\cite{Jekely2009}, heat, nutrient concentration or other chemical stimuli \cite{Moran2017}. These spatial gradients in their environment can affect the dynamics of active particles and lead to directed motion (or \textit{taxis}). Taxis can be an active response, as particles sense the local gradients and actively change their motion. Examples include \textit{E. coli} which prolongs runs when swimming up nutrient gradients to pursue nutrient rich regions \cite{Berg1972, Berg2004}. On the other hand, taxis can be a passive response, solely caused by a physical interaction with the environment that passively modulates particle dynamics. Examples of this sort include the chemotactic behavior of janus particles \cite{Baraban2013, Xiao2022} and active droplets \cite{Jin2017, Jin2018}. Inhomogeneous environments can also be leveraged to sort or organize active particles. For example, the photophobic response of \textit{E. coli} can be used to `paint' with the bacterium by subjecting a bacterial suspension to light gradients \cite{Arlt2018}. Recent research has explored imhomogeneities in the rheological properites of fluids (such as viscosity \cite{Stehnach2021, kantsler2021}, or viscoelasticity \cite{Mathijssen2016, Liu2021}) as a mechanical (as opposed to chemical or biological) mechanism of spatial control and taxis.

Spatial gradients in viscosity are prevalent in fluid environments, for example due to changes in fluid temperature or salinity. In gradients of viscosity, particles tend to perform taxis by moving up or down the gradients (defined as positive and negative viscotaxis respectively). For instance, organisms like \textit{Leptospira} and \textit{Spiroplasma} have been observed to perform positive viscotaxis \cite{kaiser1975, petrino1978, takabe2017, daniels1980} while \textit{E. coli} has been observed to perform negative viscotaxis \cite{sherman1982}. Recent experiments with \textit{Chlamydomonas rienhardtii} show contrasting behavior in weak vs strong gradients \cite{Stehnach2021}. In weak gradients, the algae accumulate in high viscosity regions due to their low speed but in strong gradients, they reorient to move towards low viscosities (negative viscotaxis).

\begin{figure}[t!]
\centering
\includegraphics[width = 0.4 \textwidth]{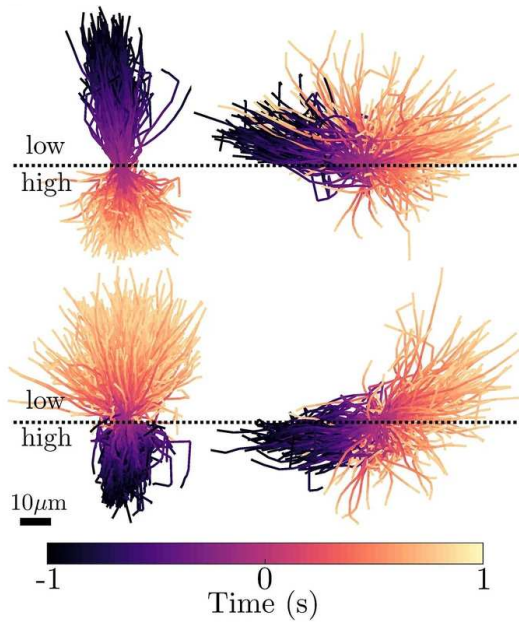}
\caption{\label{fig:experiments}Trajectories of \textit{Chlamydomonas reinhardtii} going from low to high (top) and high to low (bottom) viscosities from recent experimental work \cite{kantsler2021}. The trajectories on the left are relatively steep (closely aligned to the interface-normal) while those on the right are relatively shallow and display scattering at the interface when going from low to high viscosities. Image is from the paper by Coppola and Kanstler  \cite{kantsler2021} which is licensed under \href{https://creativecommons.org/licenses/by/4.0/}{CC BY}.}
\end{figure}

A simple fluid mechanical mechanism for viscotaxis was developed by modeling active particles as connected spheres driven by a fixed propulsive force in weak viscosity gradients \cite{liebchen2018}. Such particles exhibit taxis due to a systematic mismatch of viscous drag acting on different spheres leading to a torque that generally reorients particles to move up viscosity gradients \cite{liebchen2018}. Subsequent work modeled active particles as spherical squirmers in weak viscosity gradients, in this case the interaction of the spatially varying viscosity with the active slip boundary conditions on a squirmer generically resulted negative viscotaxis \cite{datt2019, vaseem2021}. A different swimmer, Taylors swimming sheet speeds up while moving along or against the gradients \cite{dandekar2020}. These theoretical models all rely on weak, diffuse gradients in the fluid viscosity. However recent experiments have shown very interesting particle dynamics in sharp viscosity gradients both for synthetic\cite{EsparzaLopez2021} and biological active particles \cite{kantsler2021}. In particular, we are interested here in experiments which probed the motion of \textit{Chlamydomonas reinhardtii} swimming across a sharp jump (or interface) in viscosity between miscible fluids \cite{kantsler2021}. Among other results, it was found that the algae would be quickly reoriented by the interface in viscosity and if the organism approached the interface at a sufficiently shallow angle could be reflected by the interface if going from low to high viscosity (see Fig.~\ref{fig:experiments}). Here, we develop a simple fluid dynamical model to unravel the physics underlying these experiments. We model the swimmers as spherical squirmers and focus on small, but sharp, viscosity changes. We show that the reorientation process is always in the direction of lower viscosity and derive a law, analogous to Snell's law of refraction, that governs the orientation of active particles in the presence of a viscosity interface. In analogy to ray optics, the refraction of the trajectory is always towards the medium of lower resistance. As we will show below, our theory (for pullers) matches well with experimental observations of \textit{Chlamydomonas reinhardtii} algae swimming across sharp viscosity gradients \cite{kantsler2021}. Our results are also quite similar to recent theoretical work modeling gliders moving across a substrate features a jump in frictional properties. In particular, the functional form of the reorientation law we find is identical to that found for gliders \cite{ross2021}. However that work, and other studies where the propulsive force is similarly fixed \cite{liebchen2018}, shows reorientation towards higher viscosities as one might expect due to the modulation of drag alone.

We organize the paper as follows. In the following section \ref{sec:deviation}, we provide the essential details of our model and the resulting particle dynamics. We then interpret the implications of our model and compare with experimental observations in section \ref{sec:results}. We then provide some concluding remarks and finally the technical details concerning the mathematical methods used are left to the appendix.

\section{\label{sec:deviation}A model for active particles crossing sharp viscosity gradients}

We consider an active particle immersed in an otherwise quiescent fluid, moving near and across a region where the fluid has a relative sharp change of viscosity. This change in viscosity can be due to a corresponding variation of fluid temperature, salinity, or a nutrient dissolved in the fluid. Regardless of the origin, one expects sharp viscosity gradients to vanish due to diffusion over long times, but during the short time scales over which the particle crosses the interface, relatively sharp gradients can be stable \cite{kantsler2021}. For instance, \textit{Chlamydomonas reinhardtii} algae, with a characteristic size of $\approx 10 \,\mu$m, traveling at a body length per second take $O(10\,{\rm{s}})$ to approach and cross the interface while the salinity gradients take $O(10^3\,{\rm{s}})$ to vanish \cite{kantsler2021}. We assume that the fluid viscosity is prescribed and steady, and not significantly disturbed by the presence and activity of the moving particle as in previous work \cite{liebchen2018, datt2019, vaseem2021}; however, this is an uncontrolled approximation because we assume that the sharp gradient persists, nevertheless we will show that this reasonably captures the experimental observations \cite{kantsler2021}. We note that the viscosity gradients considered here are distinct from the long-lasting viscosity differences that can exist at the interface between immiscible fluids with non-negligible surface tension that can dramatically affect (and even prevent) the particle crossing \cite{Malgaretti2016, Peter2020, Chisholm2021, Gidituri2022}.

\begin{figure}[t!]
\centering
\includegraphics[width = 0.4 \textwidth]{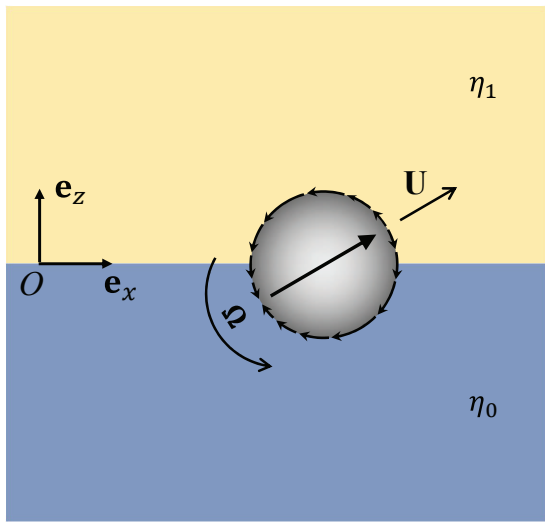}
\caption{\label{fig:schematic1}Schematic showing an active particle passing through sharp viscosity gradient and the associated coordinate system. The interface separates fluid of different viscosities $\eta_0$ and $\eta_1$. The particle radius is $a$ and its translational and rotational velocities are $\mathbf{U}$,
$\bm{\Omega}$, respectively.}
\end{figure}

For simplicity we assume changes in only one direction and choose a coordinate with the $z-$axis oriented in the direction of change, such that $\eta = \eta(z)$. The viscosity changes from one uniform viscosity $\eta(z\rightarrow -\infty) = \eta_0$ to another $ \eta(z\rightarrow\infty)=\eta_1$, and define a relative change in viscosity $\epsilon = \left(\eta_1 - \eta_0 \right)/\eta_0$. We will first assume the viscosity jumps (discontinuously) from $\eta_0$ to $\eta_1$ at $z=0$, as shown in Fig.~\ref{fig:schematic1}. In this case the viscosity field may be written
\begin{equation}
\eta(z) = \eta_0\left[1 + \varepsilon H(z)\right],
\label{eqn:shape}
\end{equation}
where $H(z)$ is the Heaviside function. This representation is of course an idealization (as finite diffusivity would instantly smooth any discontinuity); however, we will later show that relaxing this assumption to smooth changes in viscosity leaves our results unchanged. To make mathematical progress we focus on small changes in viscosity such that $\left|\varepsilon\right| \ll 1$, but we believe the main physical picture holds for any $\epsilon$.

The fluid flow generated by the active particle satisfies the incompressible Stokes equations
\begin{align}
\boldsymbol{\nabla} \cdot \bm{\sigma} = \mathbf{0}, \quad \boldsymbol{\nabla} \cdot \mathbf{u} = 0,
\end{align}
where $\mathbf{u}$ is the velocity field and stress in a Newtonian fluid, $\bm{\sigma} = - p \mathbf{I} + \eta \dot{\bm{\gamma}}$, where $p$ is the pressure, $\dot{\bm{\gamma}} = \nabla \mathbf{u} + (\nabla \mathbf{u})^T$ is the fluid strain-rate tensor.

The active particle swims with a translational velocity $\mathbf{U}$ and an angular velocity $\bm{\Omega}$ due to its activity. Thus, the velocity of the fluid on the surface of the particle $S_p$, can be decomposed as
\begin{equation}
    \mathbf{u}(\mathbf{x}\in S_p) = \mathbf{U} + \bm{\Omega} \times \mathbf{r} + \mathbf{u}^s,
\end{equation}
where $\mathbf{u}^s$ is the boundary velocity of the activity alone, $\mathbf{r} = \mathbf{x}-\mathbf{x}_c$, and $\mathbf{x}_c = \left(x_c, y_c, z_c\right)$ denotes the particle (center) position and $\dot{\mathbf{x}}_c=\mathbf{U}$. Far from the particle the fluid remains quiescent, hence
\begin{equation}
    \mathbf{u} \rightarrow \mathbf{0} \qquad \text{as} \thickspace |\mathbf{r}| \rightarrow \infty.
\end{equation}
Here we prescribe the activity of the particle $\mathbf{u}^s$, but the translational and rotational velocity are fixed by the dynamic constraints on the particle. The particle is inertialess and neutrally buoyant and with no external forcing acting on it therefore the hydrodynamic force and torque on the particle must vanish
\begin{align}
    & \mathbf{F} = \int_{S_p} \mathbf{n}_p \cdot {\bm{\sigma}} \, dS = \mathbf{0}, \\
    & \mathbf{L} = \int_{S_p} \mathbf{r} \times (\mathbf{n}_p \cdot {\bm{\sigma}}) \, dS = \mathbf{0},
\end{align}
where $\mathbf{n}_p$ is a unit normal to the particle surface. 

Specifically, we model the active particle as a spherical squirmer of radius $a$. In the squirmer model, the details of the surface activity of the swimmer are coarse-grained into a prescribed tangential slip velocity on the surface of a spherical particle \cite{lighthill1952, blake1971, ishikawa2006}. This model is particularly well-suited for ciliated microorganisms like \textit{Paramecium} and \textit{Opalina} that propel by synchronously beating numerous very small cilia on their surface, or diffusiophoretic Janus particles, which propel due to motion of a thin layer of fluid on their surface as a result of chemical gradients \cite{Moran2017}. The slip velocity is generally decomposed into Legendre polynomials (called squirming modes) in the form
\begin{equation}
    \mathbf{u}^s = \sum_{n=1}^{\infty}  \frac{2 B_n}{n(n+1)} P_n'\left(\mathbf{p} \cdot \mathbf{n}_p\right)\mathbf{p} \cdot(\mathbf{I} - \mathbf{n}_p \mathbf{n}_p ),
\end{equation}
where $\bf{p}$ is the particle orientation, $P_n$ is the Legendre polynomial of degree $n$ and $B_n$ represents the coefficients of the squirming modes. In homogeneous Newtonian fluids, the $B_1$ mode alone determines the swimming velocity (we assume here $B_1>0$), whereas $B_2$ mode is the slowest decaying contribution to the far-field flow, furthermore the second mode determines whether propulsion is primarily from the front or the back of the swimmer. Organisms, such as \emph{E. coli}, which produce propulsion from their rear end are called pushers, have $B_2 < 0$ whereas those that pull the fluid in front of them using their flagella are called pullers, such as \emph{Chlamydomonas reinhardtii}, and have $B_2 > 0$. Swimmers, with propulsion that is not distinctly fore or aft, such as \emph{Volvox carteri} with flagella uniformly distributed on its surface, are called neutral and are well described by setting $B_2 = 0$. Here we look at only the effects of the first two modes, and while one is generally only well justified in neglecting higher-order modes in the far-field, we find dynamics here to be well captured by just the $B_1$ mode.

The fluid flow field and particle velocity can be determined simultaneously by solving the Stokes equations for a force and torque free particle. Here, we take a perturbative approach; when $\epsilon \rightarrow 0$ the viscosity is uniform and the solution to a single squirmer is well known, we then obtain the leading order correction in terms of the viscosity jump $\varepsilon$, by means of the reciprocal theorem (see the reciprocal theorem subsection in Methods for more technical details).

At the leading order in $\varepsilon$, the particle moves through a homogeneous Newtonian fluid of viscosity $\eta_0$ and its velocity is well known \cite{lighthill1952, blake1971}
\begin{equation}
\mathbf{U}_0 = \frac{2}{3}B_1 \mathbf{p}, \,\, \bm{\Omega}_0 = \mathbf{0}.
\end{equation}
The viscosity variations relative to $\eta_0$ are captured at the next order and the particle rotates due to these variations at rate
\begin{equation}
    \bm{\Omega}_1 = \left( B_1 f\left( z_c \right) + B_2 g\left( z_c \right) \mathbf{n} \cdot \mathbf{p} \right) \left( \mathbf{n} \times \mathbf{p}\right),
    \label{eqn:Omega1_Final}
\end{equation}
where we have assumed a regular expansion in $\varepsilon$, $\mathbf{U}(\epsilon) = \mathbf{U}_0+\varepsilon \mathbf{U}_1+O(\varepsilon^2)$, and $\bm{\Omega}(\epsilon) = \bm{\Omega}_0+\varepsilon \bm{\Omega}_1+O(\varepsilon^2)$. Here $\mathbf{n} = \mathbf{e}_z$ is the interface normal pointing from fluid of viscosity $\eta_0$ to that of viscosity $\eta_1$ while the functions $f\left(z_c \right)$ and $g\left( z_c \right)$ depend on the particle's separation from the interface. The piecewise behavior of these functions and of the angular velocity $\bm{\Omega}_1$ is due to the fact that the particle is in contact with the viscosity interface when $\left|z_c \right| \le a$ and otherwise not. It is important to note that for $\left|z_c \right| > a$ the particle is still affected by the presence of the viscosity change due to hydrodynamic interactions mediated by the fluid at a distance.

In order to quantify particle reorientation, we simply integrate particle velocities. Noting that $\dot{\mathbf{x}}_c=\mathbf{U}$ and $\dot{\mathbf{p}}=\bm{\Omega} \times \mathbf{p}$,
we substitute the leading order results for the particle velocities and project onto the interface normal direction $\mathbf{n}=\mathbf{e}_z$ to obtain
\begin{align}
    \frac{d z_c}{dt} & = \frac{2}{3}B_1 \cos \theta + O\left(\varepsilon\right),\\
    \frac{d\theta}{dt} & = \varepsilon \left( B_1 f\left(z_c\right) + B_2 g\left(z_c\right) \cos \theta \right)\sin\theta + O\left(\varepsilon^2\right), \label{eqn:theta-time}
\end{align}
where the angle between the particle direction and the interface is defined by $\mathbf{p}\cdot\mathbf{n}=\cos\theta$. The functions $f$ and $g$ are given in the Methods section. Combining these equations gives
\begin{equation}
    \frac{d\theta}{d \medspace z_c} = \frac{3}{2}\varepsilon\bigl ( f(z_c) + \beta g(z_c) \cos\theta \bigr ) \tan\theta + O\left(\varepsilon^2 \right),
    \label{eqn:core}
\end{equation}
where $\beta = B_2/B_1$. This differential equation entirely captures the leading order effect on particle orientation $\theta$ of a viscosity jump at a distance $z_c$ from the particle center. We note that any effect of viscosity on the translational velocity of the particle would enter at $O\left(\varepsilon^2\right)$ in \eqref{eqn:core} and so is negligible compared to the leading order terms for $\epsilon \ll 1$. As we will show, equation \eqref{eqn:core} is straightforward to integrate analytically for neutral squirmers, $\beta=0$, and easily integrated numerically for pushers ($\beta < 0$) and pullers ($\beta > 0$), and the rest of this paper are results and discussion that arise out of it.

\section{Results}\label{sec:results}
We begin first with analytical results for neutral squirmers, $\beta=0$, before proceeding to present numerical results for pushers ($\beta < 0$) and pullers ($\beta > 0$) and a comparison to recent experiments for pullers.

\begin{figure}[t!]
\subfloat{\includegraphics[scale = 0.47]{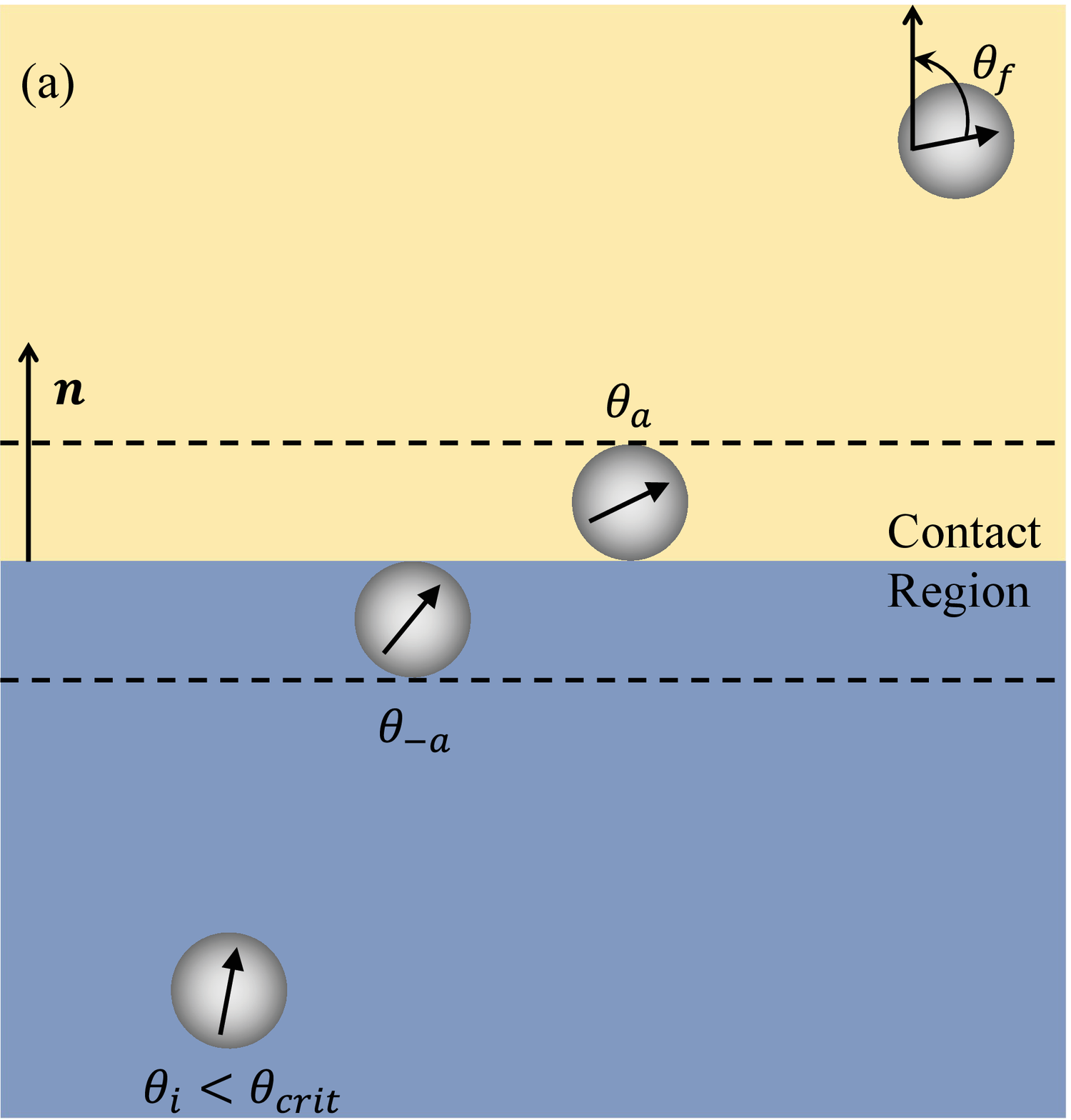}\label{fig:fig3a}} \hfill
\subfloat{\includegraphics[scale = 0.47]{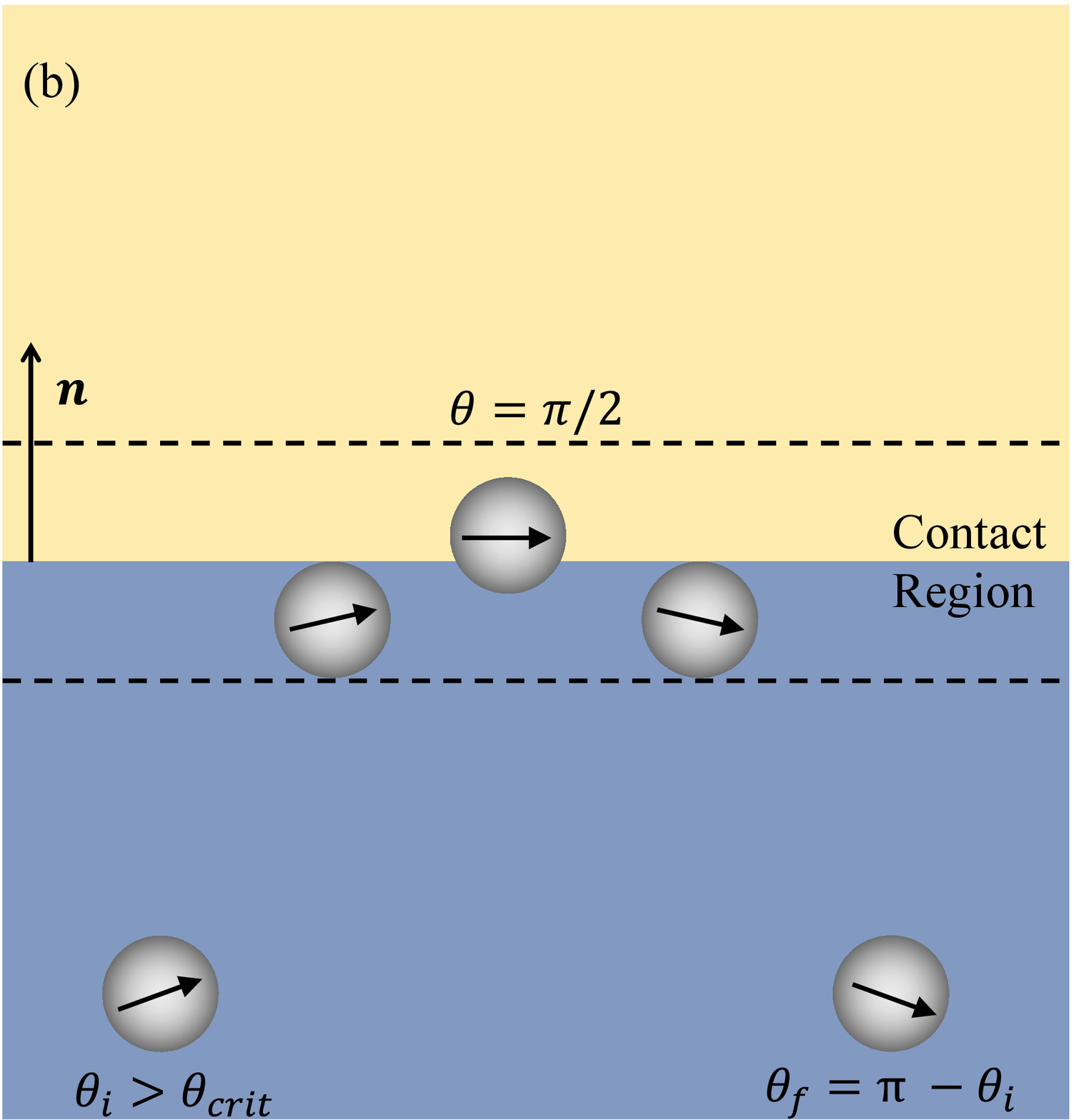}\label{fig:fig3b}}
\caption{\label{fig:schematic2} Schematic showing the reorientation of an active particle as it ($\bf{a}$) crosses a viscosity interface  or ($\bf{b}$) gets reflected by the interface. This reorientation depends largely on the viscosity difference $\eta_1-\eta_0$ and the particle is reflected only when going from low to high viscosity if its initial orientation is sufficiently shallow, $\theta_i > \theta_{crit}$.}
\end{figure}

\subsection{\label{sec:neutral}Neutral squirmers}
The reorientation of a neutral swimmer can be immediately understood by examining the instantaneous rotational dynamics. To leading order equation \eqref{eqn:theta-time} with $B_2=0$  simplifies to
\begin{equation}
    \frac{d \medspace \theta}{d \medspace t} = \varepsilon B_1 f(z_c) \sin \theta.
    \label{eqn:qualitative equation}
\end{equation}
Noting that $B_1$ and $f(z_c)$ are both positive, when $\varepsilon > 0$ we see that $\theta=0$ is an unstable fixed point and all orientations flow towards $\theta =\pm\pi$. Conversely, for $\varepsilon < 0$ all orientations flow to $\theta=0$. This means that no matter the orientation or position, the particle is always reorienting to align along $\mathbf{n} = \mathbf{e}_z$ and point in the direction of the lower viscosity, consistent with results for squirmers in weak viscosity gradients \cite{datt2019, vaseem2021}. Because $f\propto z_c^{-4}$, the reorientation rate decreases very quickly with distance from the interface, and thus the reorientation process is ultimately dominated by contact with the interface. One consequence of these dynamics is that a particle going from low to high viscosity can be scattered off the interface depending on its incident orientation.

To quantify the reorientation we note that \eqref{eqn:core} is separable when $\beta=0$, integrating we obtain
\begin{align}
\frac{\sin\theta_f}{\sin\theta_i} = \exp\left[\frac{3}{2}\epsilon\int_{z_i}^{z_f}f(z_c)dz_c\right],
\label{eqn:integration}
\end{align}
where $\theta_i$ is the orientation at an initial position $z_i$ and likewise $\theta_f$ is the final orientation at $z_f$. 

We define the `total' reorientation caused by the interface as the particle crosses from far on one side to far on the other to be the limit when $z_i\rightarrow -\infty$ and $z_f\rightarrow \infty$ (when the particle goes from $\eta_0$ to $\eta_1$). In this case the integral simply equals $1/3$ and the total reorientation is given by the formula
\begin{equation}
    \sin \theta_{f} = \exp\left[\frac{\eta_1-\eta_0}{2\eta_0}\right] \sin \theta_{i}.
    \label{eqn:SnellsFinal}
\end{equation}
This formula bears striking similarity to Snell's law of refraction, except here the `relative refractive index' is given by the exponentiated relative viscosity difference. The reorientation is independent of the speed of the particle due to the linearity of the Stokes equations, in this case both the thrust generated by the particle and the drag felt by the particle would be proportional to the $B_1$ mode. This form of reorientation law, $\sin\theta_f=e^\alpha\sin\theta_i$, was found for gliders moving across a substrate featuring a jump in frictional properties \cite{ross2021}. In that case $\alpha = -2a\zeta_{rt}/\zeta_{rr}$ where $\zeta_{rr}$, $\zeta_{rt}$ are torque-rotation and torque-translation resistance coefficients of the particle respectively. The similarities arise because in both cases the particles are subject to linear drag laws.

Unlike the refraction of light, or gliders on a substrate, squirmers interact (hydrodynamically) with the interface from any point in space, but the functional form of the interaction, given by $f(z_c)$ changes upon contact. Because of this we integrate equation \eqref{eqn:integration} in multiple stages, separately accounting for the particle's approach to the interface $(z_c = -\infty \to -a$, $\theta = \theta_i \to \theta_{-a})$, crossing the interface $\left(z_c = -a \to +a, \theta = \theta_{-a} \to \theta_a \right)$ and the departure from the interface $(z_c = +a \to +\infty$, $\theta = \theta_a \to \theta_f)$ as illustrated in Fig.~\ref{fig:fig3a}. In this way we can quantify the starting $\left(\theta_{start} \right)$ and ending $\left( \theta_{end}\right)$ orientation in each of these stages
\begin{equation}
    \sin \theta_{end} = e^{\alpha} \sin \theta_{start},
    \label{eqn:Snell's law for NS}
\end{equation}
where for $\theta_{start} = \left \{ \theta_i, \theta_{-a}, \theta_a \right \}$, and $\theta_{end} = \left \{ \theta_{-a}, \theta_a, \theta_f\right \}$, we find $\alpha = \left \{ \frac{\varepsilon}{32}, \frac{7 \varepsilon}{16}, \frac{\varepsilon}{32}\right \}$, in that order in each of the three stages. We can also relate the particle's initial and final orientations by combining equation \eqref{eqn:Snell's law for NS} in all three stages. We notice that the amount of reorientation during the approach and departure from the interface is the same. However, the large value of $\alpha$ means that the reorientation process is dominated during contact with the interface $\left(\theta_{-a} \to \theta_a\right)$. And, as discussed earlier, the reorientation process is always in the direction of lower viscosity. In analogy to ray optics, the refraction of the trajectory is always towards the medium of lower resistance. A similar preference was also shown by active particles in linear or diffuse viscosity gradients \cite{datt2019, Stehnach2021, vaseem2021}, and as we will show below (for pullers), matches well with experimental observations of \textit{Chlamydomonas reinhardtii} algae swimming across sharp viscosity gradients \cite{kantsler2021}. In contrast, studies done where the propulsive force is fixed, both for swimmers in diffuse viscosity gradients and gliders across a frictional substrate, show reorientation towards higher viscosities as one might expect due purely to the modulation of drag.

We note that in deriving \eqref{eqn:SnellsFinal} we assume that the trajectory from one side to the other is physically realizable. This is not always the case. If the active particle is swimming towards a higher viscosity $\eta_1 > \eta_0$, with a sufficiently shallow angle it may be reoriented back (reflected by the interface). But note, due to hydrodynamic interactions, the particle may be reoriented back before even coming into contact with the interface, or even after completely crossing the interface. To examine this phenomena we find the limit of validity of \eqref{eqn:SnellsFinal}, which we define $\theta_i=\theta_{crit}$ which occurs when $\theta_f = \pi/2$, that is the swimmer is tangent to the interface at $z_f\rightarrow \infty$, this case we find
\begin{equation}
\theta_{crit} = \arcsin\left(\exp\left[\frac{\eta_0-\eta_1}{2\eta_0}\right]\right).
\label{eqn:thetacrit}
\end{equation}
Therefore, for $\eta_1>\eta_0$ equation \eqref{eqn:SnellsFinal} is valid only when $\theta_i \le \theta_{crit}$, as a particle with an initial angle, $\theta_i > \theta_{crit}$ (a sufficiently shallow angle of approach to the interface), will be reflected back (it will not reach $z_f\rightarrow \infty$). We can likewise define critical initial angles such that the particle does not cross the interface  $\theta_{crit,a} = \arcsin\left(\exp\left[15(\eta_0-\eta_1)/(32\eta_0)\right]\right)$, or even touch the interface $\theta_{crit,-a} = \arcsin\left(\exp\left[(\eta_0-\eta_1)/(32\eta_0)\right]\right)$, using \eqref{eqn:Snell's law for NS}. We see that, much like the total reorientation, $\theta_{crit}$ in \eqref{eqn:thetacrit} is dominated by particles which are scattered at the interface, with only a narrow set of initial angles that lead to reflection before or after contact with the interface. Regardless of where the particle is reflected, the entire scattering process is symmetric (about the point when $\theta=\pi/2$ as shown in Fig.~\ref{fig:fig3b}) and hence obeys the reflection law
\begin{align}
\theta_f = \pi - \theta_i \quad 
\end{align}
for all particles when $\theta_i > \theta_{crit}$, as previously shown for gliders \cite{ross2021}.

We have thus far assumed a physically unrealistic discontinuous viscosity change and a simple neutral squirmer in order to derive these simple formulas. In the following sections we relax these assumptions and find that neither significantly impact the reorientation process.

\subsection{Smooth gradients}
To investigate the reorientation and scattering of the particle due to a viscosity change that varies smoothly (due to the effects of diffusion), instead of a Heaviside function in \eqref{eqn:shape} we say $H(z)=\left(1+\tanh(kz)\right)/2$ where $k > 0$ and $1/k$ is the effective length scale over which the viscosity varies between $\eta_0$ and $\eta_1$. Hence, this viscosity variation approaches the discontinuous profile used previously as $k \to \infty$. Calculation proceeds similarly to that in the sharp gradients. The angular velocity required for this calculation is found by substituting the $\tanh$ viscosity profile in the reciprocal theorem \eqref{eqn:Omega1} and its expression looks the same as that found in sharp viscosity gradients \eqref{eqn:Omega1_Final} except for the functions $f\left(z_c \right)$ and $g\left(z_c\right)$ which are given in the methods section.

Despite the differences in viscosity profile and angular velocity, we find no difference in the overall reorientation $\left( \theta_i \to \theta_f\right)$ between smooth and sharp viscosity gradients. The law governing the reorientation in smooth gradients is identical to that found in sharp gradients \eqref{eqn:SnellsFinal}. This implies that the critical orientation required for scattering in smooth and sharp viscosity gradients is also the same. It appears that, as with the refraction of light, the interface between fluids of differing viscosity can be smoothed out and the total reorientation remains unchanged. This is surprising, because unlike light, the active particle interacts non-locally with the entire medium at once at all times due to hydrodynamics.

\subsection{Pushers and pullers}
\begin{figure}[t!]
\subfloat{\includegraphics[scale = 0.47]{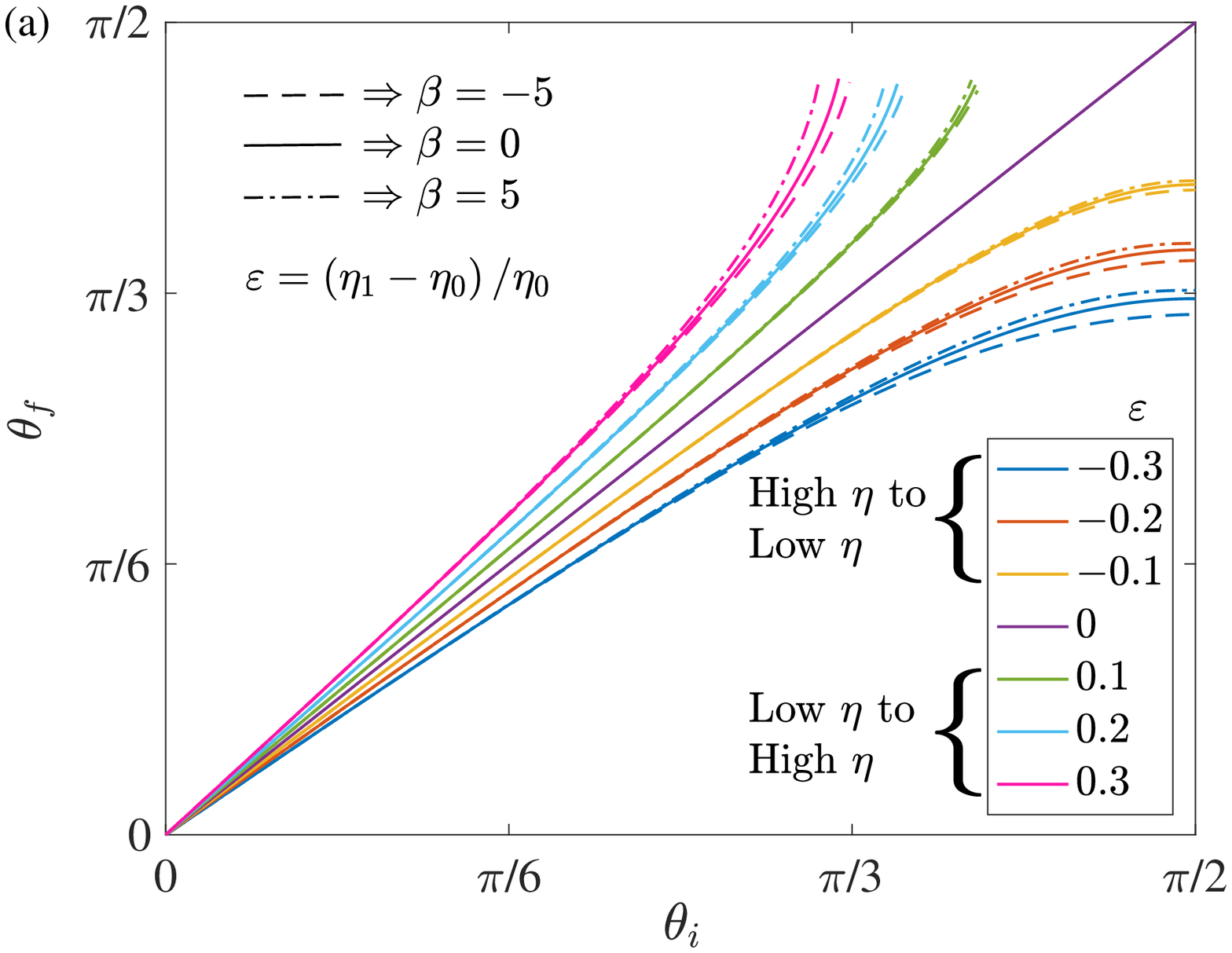}\label{fig:fig4a}} \hfill
\subfloat{\includegraphics[scale = 0.47]{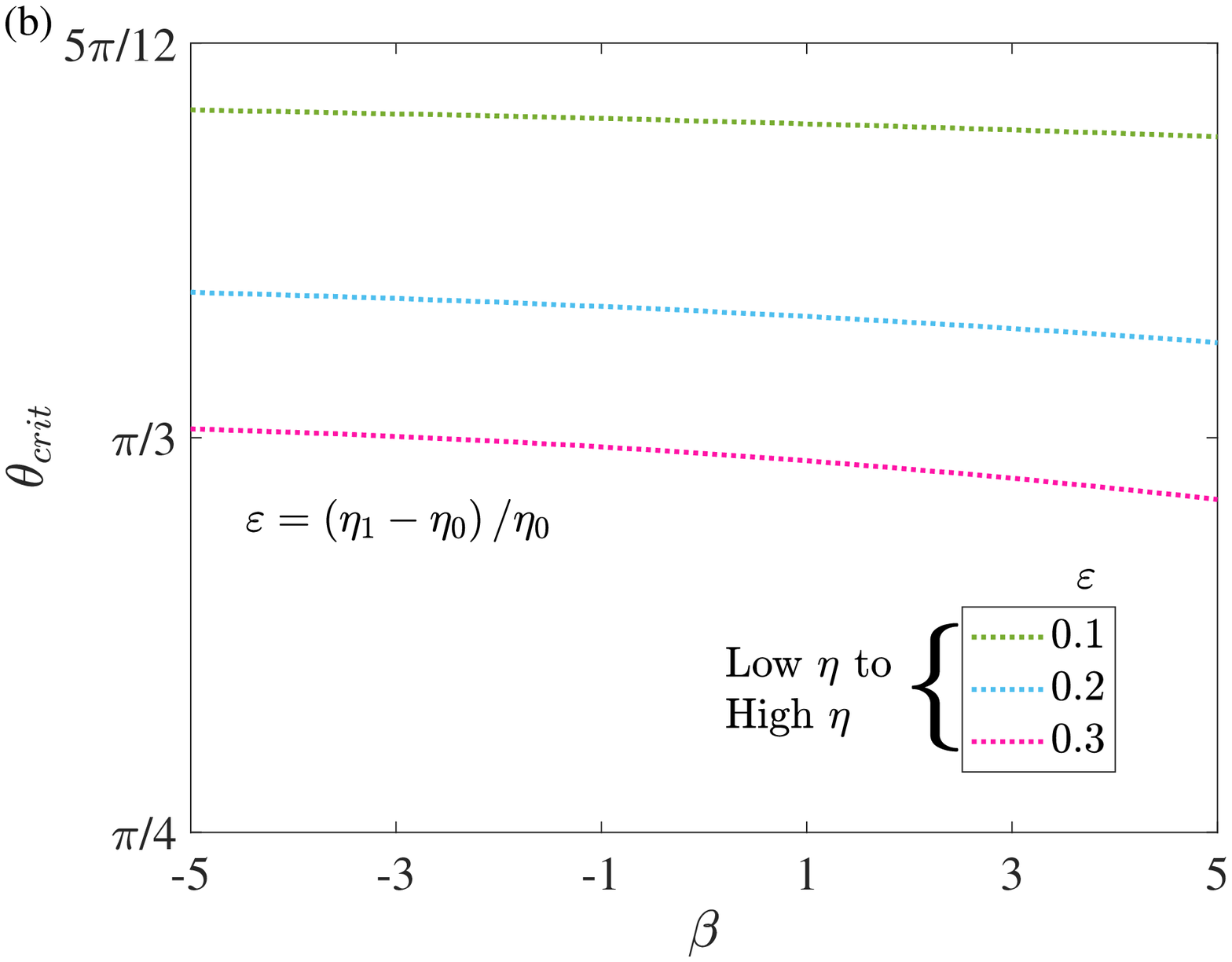}\label{fig:fig4b}}
\caption{ The reorientation of active particles that cross the interface ($\bf{a}$) and the critical orientation required for scattering from the interface ($\bf{b}$). Here, the solid, dashed, and dash-dotted lines in ($\bf{a}$) correspond to the neutral swimmers $\beta = 0$, pushers $\beta = -5$, and pullers $\beta = 5$, respectively. On the other hand, the different line colors in ($\bf{a}$),($\bf{b}$) represent different viscosity jumps $\varepsilon$.}
\end{figure}

For pushers and pullers the differential equation governing the reorientation \eqref{eqn:core} is not separable and so we compute the reorientation numerically for $\beta\ne0$. We find that the reorientation and scattering of pushers or pullers are similar to those of neutral swimmers with only a weak dependence on the squirming ratio $\beta$. See Fig.~\ref{fig:fig4a} for the reorientation of the swimmers crossing the interface and Fig.~\ref{fig:fig4b} for the critical orientation required for scattering, obtained from the numerical solution of \eqref{eqn:core}. Hence, the pushers and pullers, like the neutral swimmers, orient towards regions of lower viscosity. Going from high to low viscosity ($\varepsilon < 0$) pullers rotate slightly less while the pushers rotate slightly more than the neutral swimmer. Conversely going from low viscosity to high viscosity ($\varepsilon > 0$) pullers rotate more while the pushers rotate less than a neutral swimmer thereby very weakly changing $\theta_{crit}$ as shown in Fig.~\ref{fig:fig4b}.

Part of the reason for the weak dependence of reorientation on the $B_2$ mode occurs because the rotation caused by this mode before and after crossing the interface are in the opposite direction (as $g\left(z_c\right)$ is an odd function) and hence counteract each other (they do not cancel due to the cosine term in \eqref{eqn:core}). Conversely, as discussed previously the rotation caused by the $B_1$ mode is always in the same direction before and after crossing the interface (as $f\left(z_c\right)$ is an even function).

The weak dependence of reorientation on $\beta$ can be leveraged to find the reorientation experienced by the pushers and pullers analytically. This is achieved by expanding the leading order (in $\varepsilon$) orientation $\theta$ in terms of $\beta$ and solving equation \eqref{eqn:core} at each order in $\beta$ as shown in the Methods section. In principle, such a perturbation holds for only $\left|\varepsilon\right|\ll \left| \beta \right| \ll 1$ but the weak functional dependence yields accurate results up to $\left|\beta \right| \approx 10$ for any $\left| \varepsilon \right| \ll 1$.

\subsection{Comparison to experiment}

\begin{figure}[t!]
\centering 
\subfloat{\includegraphics[scale = 0.32]{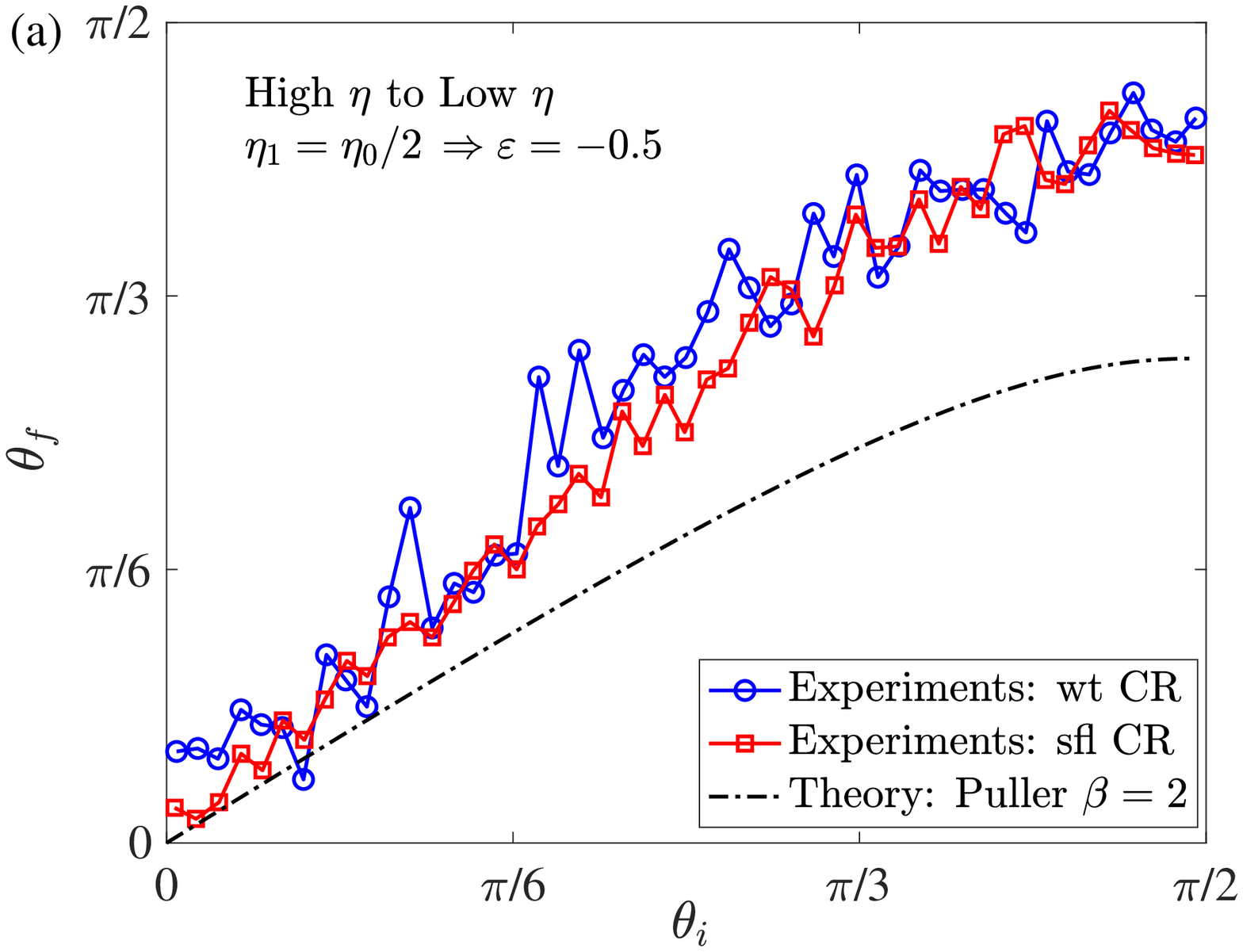}\label{fig:fig5a}}
\subfloat{\includegraphics[scale = 0.32]{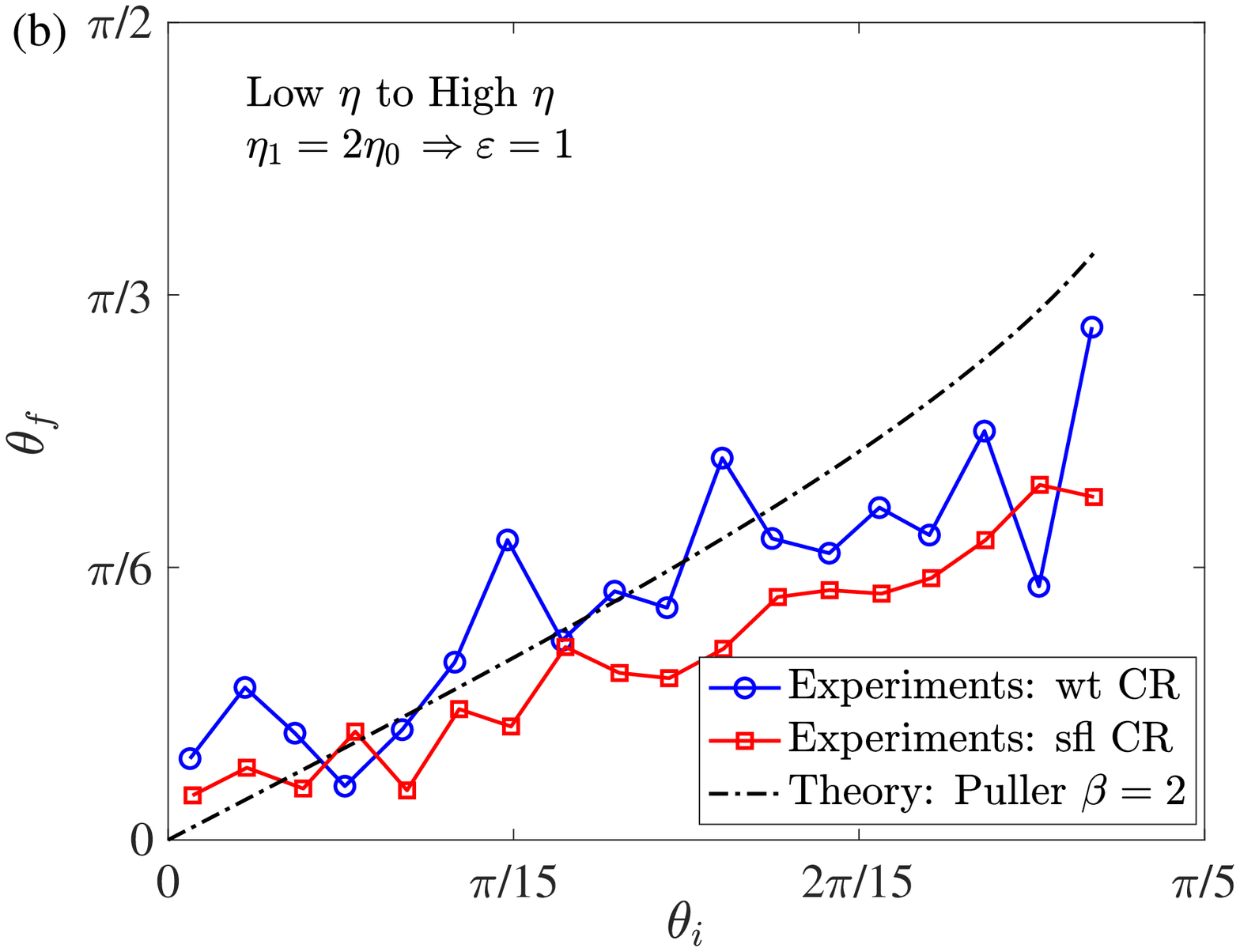}\label{fig:fig5b}}
\subfloat{\includegraphics[scale = 0.32]{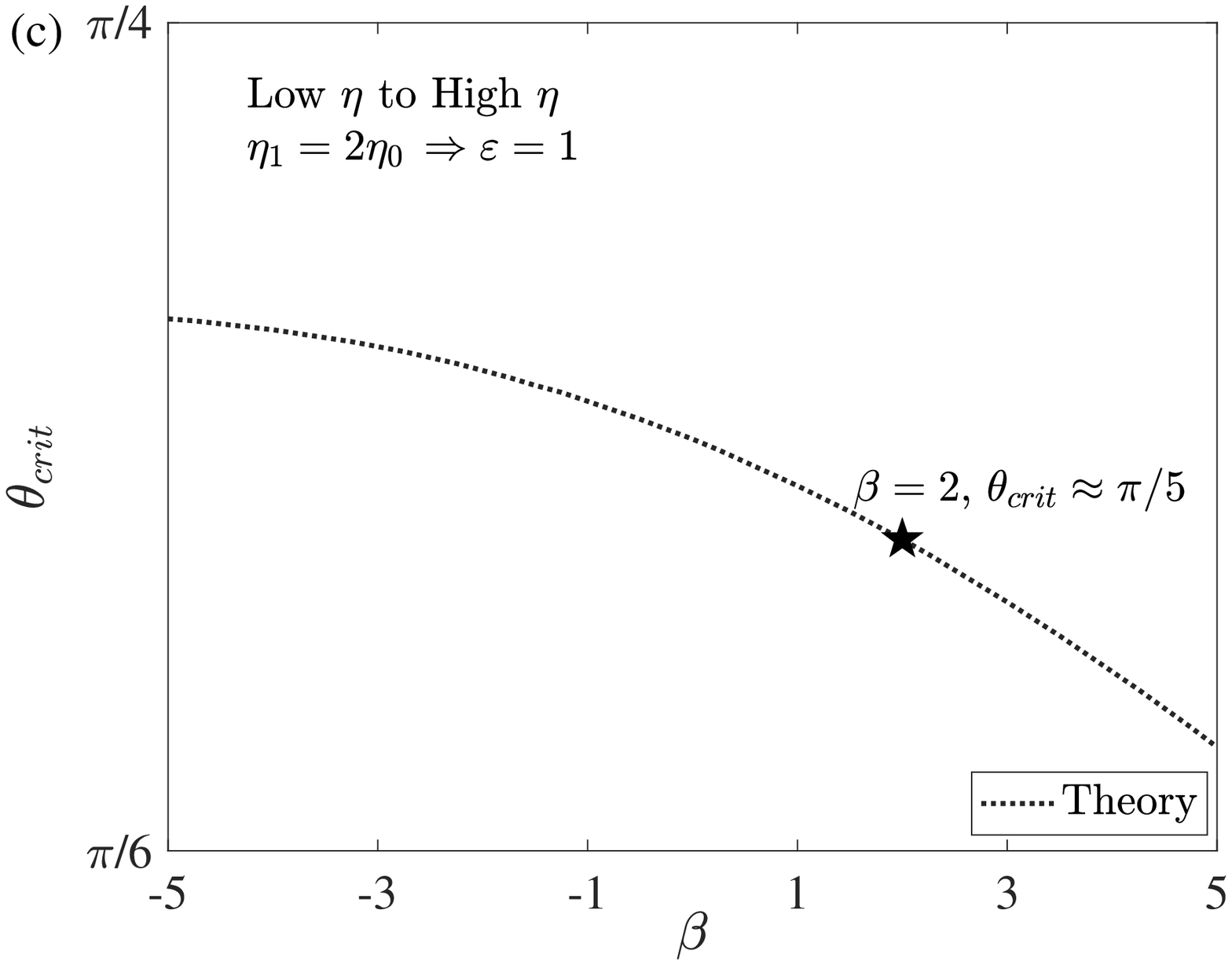}\label{fig:fig5c}}
\caption{\label{fig:comparsion}The reorientation of active particles that cross the interface ($\bf{a}$), ($\bf{b}$), and the critical orientation required for scattering from the interface ($\bf{c}$). Here, the lines with symbols in ($\bf{a}$), ($\bf{b}$) represent the previous experiments using wild-type (wt) or short-flagellated (sfl) {\textit{Chlamydomonas reinhardtii}} \cite{kantsler2021} while the dash-dotted lines correspond to the current theory for pullers $\beta = 2$. Also, the only line in ($\bf{c}$) represents the current theory.}
\end{figure}
We now compare our theory with recent experiments conducted with both wild-type (wt) and short-flagellated (sfl) {\textit{Chlamydomonas reinhardtii}} (CR), swimming across sharp viscosity gradients \cite{kantsler2021}. Just as we have predicted above, the CR were found generically to reorient towards lower viscosities and there was a critical angle, going from low to high viscosity, past which the swimmers would be reflected by the viscosity interface. The initial and final orientations of the algae were recorded $1s$ before reaching and $1s$ after crossing the interface in the experiments. Using experimental velocities this equates to $z_i \approx -3a$ and $z_f \approx 3a$ where $a = 5 \mu$m is the approximate swimmer radius, and we use these values in our theory for comparison. In the experiments, the viscosity of one fluid (water) was held constant ($\eta = 10^{-3}$Pa.s) and a variety of different viscosities were used for the other fluid from $2-62\times$ greater by dissolving varying concentrations of methylcellulose in water, resulting in relative viscosity differences  $\left| \varepsilon\right| = 0.5-61$. We compare our asymptotic theory, which assumes $\varepsilon \ll 1$, only to the smallest values $\varepsilon = -0.5, 1$ representing particle motion from high to low and low to high viscosities respectively. We note that the experimental data indicates small but systematic reorientation even in homogeneous Newtonian fluid `control' experiments. In order to remove this effect we subtracted the reorientation reported in homogeneous fluids from that in finite viscosity gradients and compared the difference with the theory. Lastly, we estimated the squirming ratio $\beta = 2$ by taking the $B_1$ value from the known swimming velocity in a homogeneous fluids $ \approx 100\,\mu \text{m/s}= \frac{2}{3}B_1$ and $B_2$ value from the stresslet exerted by the CR, $10\,{\rm{pN}} \times 10\,\mu{\rm{m}} \approx 4\pi\eta a^2 B_2$ found in other experiments \cite{Goldstein2015}. In this parameter regime, our theoretical model matches experimental observations well. In Fig.~\ref{fig:fig5a} we show the reorientation ($\theta_f$ vs $\theta_i$) for swimming from high to low viscosity ($\epsilon=-0.5$) while Fig.~\ref{fig:fig5b} shows swimming from low to high viscosity ($\epsilon=1$). In both cases our model somewhat over-predicts the amount of reorientation but captures nicely the general qualitative features observed in experiment. Over-predicting the reorientation naturally leads to a critical angle $\left( \theta_{crit} \approx \pi/5 \right)$ found in theory that is lower than that reported in the experiments $\left(\theta_{crit} = \pi/3 \right)$, where a shallower approach to the interface is needed to scatter (see Fig.~\ref{fig:fig5c}). Quantitative differences are not surprising as our perturbative approach assumes $\varepsilon \ll 1$ while in experiments at best we have $\varepsilon = O(1)$ . Another possible cause of quantitative discrepancy may be due to confinement of the algae, in a microfluidic channel of height 20 $\mu$m, in the experiments unlike the free-space assumption made in the theory. Finally we assumed that the swimmer does not stir the viscosity field due to its motion and any mixing of the fluid in experiments is likely to weaken the effect of viscosity differences on reorientation.

\section{Conclusions}

Motivated by the recent experiments showing {\textit{Chlamydomonas reinhardtii}} algae scattering at sharp viscosity gradients \cite{kantsler2021}, we developed a simple analytical model for active particles swimming across sharp changes in the viscosity of the suspending fluid. We found that pushers, pullers and neutral swimmers all interact similarly with the interface. Swimmers are generically reoriented towards the region of lower viscosity (as found in previous studies with weak gradients \cite{datt2019, vaseem2021}). As a result, if active particles approach a viscosity interface at a sufficiently shallow angle they can be reflected if swimming from low to high viscosity; otherwise, they simply cross the interface undergoing a degree of reorientation set by the relative viscosity difference. This is similar to the refraction or reflection of the light due to a change in refractive index and the the law we derive governing the reorientation of neutral swimmers similar to Snell's law of ray optics (as previously shown for gliders on a frictional substrate \cite{ross2021}). Our theory compares very well with experimental observations \cite{kantsler2021} and provides a simple model for the dynamics of active particles in fluids with inhomogeneous viscosity. These results suggest that tailoring the mechanical properties of fluids can be an effective method to sort and control active matter.

\section*{Acknowledgements}
The authors gratefully acknowledge funding (RGPIN-2020-04850) from the Natural Sciences and Engineering Research Council of Canada (NSERC).

\appendix

\section{\label{sec:The Reciprocal Theorem} Reciprocal Theorem}

The dynamics of a force-free and torque-free active particle in a fluid medium of arbitrary rheology is given by \cite{Elfring2017}
\begin{equation}
       \pmb{\mathsfit{U}} = \hat{\pmb{\mathsfit{R}}}_{\pmb{\mathsfit{FU}}}^{-1} \cdot (\pmb{\mathsfit{F}}_s + \pmb{\mathsfit{F}}_{NN}),
       \label{eqn:stokes}
\end{equation}
where $\pmb{\mathsfit{U}}=[\mathbf{U} \thickspace \bm{\Omega} ]^\mathsf{T}$ is a six-dimensional vector containing the swimmer's translational and angular velocities, and likewise $\pmb{\mathsfit{F}} = [\mathbf{F} \thinspace \thickspace \mathbf{L} ]^\mathsf{T}$ contains force and torque. $\hat{\pmb{\mathsfit{R}}}_{\pmb{\mathsfit{FU}}}$ is the resistance tensor for the particle in a fluid of uniform viscosity $\eta_0$, $\pmb{\mathsfit{F}}_s$ is the thrust force and torque due to particle activity in a homogeneous Newtonian fluid of viscosity $\eta_0$, while the additional force $\pmb{\mathsfit{F}}_{NN}$ accounts for the changes in the rheological properties (viscosity) of the fluid. The formulas are obtained by using the reciprocal theorem, by projecting onto a known auxiliary flow as an adjoint solution (denoted by a hat),
\begin{align}
      \pmb{\mathsfit{F}}_s &= \int_{S_p} \mathbf{u}^s \cdot (\mathbf{n}_p \cdot \hat{\pmb{\mathsfit{T}}}_{\pmb{\mathsfit{U}}}) \thickspace dS,\\
      \pmb{\mathsfit{F}}_{NN} &= - \int_{\mathcal{V}} \pmb{\tau}_{NN} : \hat{\pmb{E}}_{\pmb{\mathsfit{U}}} \thickspace dV,  
\end{align}
where $\mathcal{V}$ denotes the entire fluid volume outside the particle and $\pmb{\tau}_{NN}= \boldsymbol{\sigma} +p\mathbf{I}-\eta_0\dot{\boldsymbol{\gamma}} = (\eta-\eta_0)\dot{\boldsymbol{\gamma}}$ represents the extra (deviatoric) stress due to changes in viscosity from $\eta_0$. $\hat{\pmb{\mathsfit{T}}}_{\pmb{\mathsfit{U}}}$, $\hat{\pmb{E}}_{\pmb{\mathsfit{U}}}$ are linear operators relating the stress and strain-rate in the fluid to particle velocity, $\hat{\pmb{\sigma}} =
\hat{\pmb{\mathsfit{T}}}_{\pmb{\mathsfit{U}}} \cdot \hat{\pmb{\mathsfit{U}}} $ and
$\hat{\dot{\pmb{\gamma}}} = 2 \hat{\pmb{E}}_{\pmb{\mathsfit{U}}} \cdot \hat{\pmb{\mathsfit{U}}}$ in a fluid of homogeneous viscosity $\eta_0$.

To facilitate the evaluation of swimming velocity in equation \eqref{eqn:stokes}, we assume small relative viscosity differences $\varepsilon \ll 1$ and regular perturbation expansion for any functional dependence on $\varepsilon$, for example $h\left(\varepsilon\right) = h_0 + \varepsilon h_1 + \ldots$. In this way, because the extra stress due to viscosity changes is $O\left(\varepsilon\right)$,
\begin{align}
\bm{\tau}_{NN} = \left(\eta - \eta_0\right) \dot{\bm{\gamma}} = \varepsilon \eta_0 H\left(z\right) \dot{\bm{\gamma}} \sim O\left(\varepsilon\right),
\end{align}
the swimmer is moving through a homogeneous Newtonian fluid of viscosity $\eta_0$ to leading order and its velocity is well known
\begin{align}
    \label{eqn:tran_vel_leading_order}
    \mathbf{U}_0 & = \frac{1}{6 \pi \eta_0 a} \int_{S_p} \mathbf{u}^s \cdot (\mathbf{n}_p \cdot \hat{\pmb{\mathsfit{T}}}_{\mathbf{U}}) \thickspace dS  = \frac{2}{3} B_1 \mathbf{p} , \\
    \bm{\Omega}_0 & = \frac{1}{8 \pi \eta_0 a^3} \int_{S_p} \mathbf{u}^s \cdot (\mathbf{n}_p \cdot \hat{\pmb{\mathsfit{T}}}_{\bm{\Omega}}) \thickspace dS  = \bm{0}.
\end{align}
The effects of viscosity variations relative to $\eta_0$ are captured at the next order, where the swimming velocity is
\begin{align}
    \mathbf{U}_1 & = - \frac{1}{6 \pi \eta_0 a} \int_{\mathcal{V}} \bm{\tau}_{NN,1} :
    \hat{\pmb{E}}_{\mathbf{U}}\thickspace dV , \label{eqn:U1}\\
    \bm{\Omega}_1 & = - \frac{1}{8 \pi \eta_0 a^3} \int_{\mathcal{V}} \bm{\tau}_{NN,1} :\hat{\pmb{E}}_{\bm{\Omega}}\thickspace dV . \label{eqn:Omega1}
\end{align}
Here, $\bm{\tau}_{NN,1} = \eta_0 H\left(z\right) \dot{\bm{\gamma}}_0$ and $\dot{\bm{\gamma}}_0 = \nabla \mathbf{u}_0 + (\nabla \mathbf{u}_0)^T$ is the rate of strain tensor associated with the leading order flow $\mathbf{u}_0$. These small viscosity variations $\varepsilon \eta_0 H\left(z\right)$ alter the velocity of swimmer in homogeneous fluid $\mathbf{U}_0$, $\bm{\Omega}_0$ by a small correction $\mathbf{U}_1$, $\bm{\Omega}_1$. An evaluation of integrals in equations \eqref{eqn:U1}, \eqref{eqn:Omega1} with discontinuous visocisty jump ($H$ is a Heaviside function) yields this correction as
\begin{align}
    \mathbf{U}_1 &= B_1 \left( A(z_c) \mathbf{n} + B(z_c) \mathbf{p} \right) + B_2 \left( C(z_c) \mathbf{n} + D (z_c) (\mathbf{qq} \cdot \mathbf{n}) + E (z_c) (\mathbf{n} \cdot \mathbf{p})^2 \mathbf{n} \right)  ,     \label{eqn:tran_vel_first_order}\\
    \bm{\Omega}_1  &= \bigl (B_1 f(z_c) + B_2 g(z_c) (\mathbf{n} \cdot \mathbf{p})  \bigr ) (\mathbf{n} \times \mathbf{p}), \label{eqn:rot_vel_first_order}
\end{align}
where for $\left|z_c\right|>a$
\begin{gather}
        A\left(z_c\right) = \frac{a^3(-a^2 + 3 z_c^2)}{24 z_c^{5}},\thickspace B\left(z_c\right) = \frac{a^3(-a^2 + z_c^2)}{24 z_c^{5}},\thickspace C\left(z_c\right) = \frac{a^2(-5a^4 + 12 a^2 z_c^2 - 9 z_c^4) }{96 z_c^{6}},\nonumber\\
        D\left(z_c\right) = \frac{a^2(5a^4 -9 a^2 z_c^2 + 9 z_c^4) }{48 z_c^{6}},\thickspace E\left(z_c\right) = \frac{a^2(5a^4 - 18 12 a^2 z_c^2 + 9 z_c^4) }{96 z_c^{6}},\\
        f\left(z_c\right) = \frac{a^3}{16 z_c^4},\thickspace g\left(z_c\right) = \frac{(-4a^2 + 3z_c^2)a^2}{32 z_c^5},\label{eqn:coefffaraway}
\end{gather}
and for $\left|z_c\right|\le a$
\begin{gather}
        A\left(z_c\right) = \frac{ z_c(3a^2 - z_c^2)}{24 a^3},\thickspace B\left(z_c\right) = \frac{ z_c(a^2 - z_c^2)}{24 a^3},\thickspace C\left(z_c\right) = \frac{-5a^4 + 6 a^2 z_c^2 - 3 z_c^4}{96a^4},\nonumber \\
        D\left(z_c\right) = \frac{8 a^4 -3 z_c^4}{48 a^4},\thickspace E\left(z_c\right) = \frac{-a^4 - 18 a^2 z_c^2 + 15 z_c^4}{96a^4},\\
        f\left(z_c\right) = \frac{3a^2 - 2 z_c^2}{16 a^3},\thickspace g\left(z_c\right) = \frac{(-7a^2 + 6 z_c^2)z_c}{32 a^4}.\label{eqn:coeffcrossing}
\end{gather}

When we assume a smooth viscosity profile $H(z)=\left(1+\tanh(kz)\right)/2$, we obtain
\begin{align}
    f (z_c) & = \int_{-\infty}^{-a} \frac{a^3 \{ 1+ \tanh [k(z+z_c)] \}}{8z^5} d z + \int_{-a}^{a} \frac{z \{ 1+ \tanh [k(z+z_c)] \}}{8a^3} d z +  \int_{a}^{\infty} \frac{a^3 \{ 1+ \tanh [k(z+z_c)] \}}{8z^5} d z, \\
    g (z_c) & = \int_{-\infty}^{-a} \frac{(20a^4 - 9a^2z^2) \{ 1+ \tanh [k(z+z_c)] \}}{64z^6} d z  + \int_{-a}^{a} \frac{(18z^2 -7 a^2) \{ 1+ \tanh [k(z+z_c)] \}}{64a^4} d z \nonumber \\
   & \quad  +  \int_{-\infty}^{-a} \frac{(20a^4 - 9a^2z^2) \{ 1+ \tanh [k(z+z_c)] \}}{64z^6} d z.
\end{align}
\color{black}

\section{\label{sec:app_perturbation}Pushers and pullers, the effect of $\beta$}
In order to find the leading order effect of the second squirming mode for pushers and pullers, we assume $\beta\ll 1$ and  perform a regular perturbation expansion of the orientation in $\beta$
\begin{equation}
    \theta = \theta_0 + \theta_1 \beta + O(\beta^2, \varepsilon).
    \label{eqn:analytical assumption}
\end{equation}
Here, we assumed $\left| \beta \right| \gg \left| \varepsilon\right|$ and retained the terms at $O\left( \beta\right)$ unlike those at $O\left( \varepsilon\right)$. We substitute this expansion in \eqref{eqn:core} and solve the resulting equation at each order in $\beta$. At zeroth order, pushers or pullers become neutral swimmers and the orientation is given by \eqref{eqn:SnellsFinal}. Any deviations relative to the reorientation of the neutral swimmer are captured at the next order where
\begin{equation}
    \frac{d \medspace \theta_1}{d \medspace z_c} - \frac{3\varepsilon}{2} \frac{f(z_c)}{\cos^2 \theta_0} \theta_1
    = \frac{3}{2} \varepsilon g(z_c) \sin \theta_0.
    \label{eqn:first order analytical result}
\end{equation}
The initial condition is $\theta_1 = 0$ as $z_c \to -\infty$. We solve \eqref{eqn:first order analytical result} in multiple stages, separately accounting for the reorientation during the interface approach, crossing and departure. We find
\begin{align}
    \label{eqn:analytical approaching}
    \left. \theta_1 \right|_{z_c = -a} & = \frac{G_1}{\sqrt{e^{-\frac{\varepsilon}{16}} - \sin^2 \theta_i}}, \\
    G_1 & = \int_{-\infty}^{-1} \varepsilon \sin \theta_i \sqrt{1 - e^{\frac{\varepsilon}{16 z^3}} \sin^2 \theta_i}
    \frac{\left(- 12 + 9 z^2\right)}{64 z^5} d z. \nonumber
\end{align}
as the particle touches the interface. Then
\begin{align}
        \label{eqn:analytical crossing}
       \left. \theta_1 \right|_{z_c = a} & = \frac{ G_2 + G_1 }{\sqrt{e^{\frac{-15\varepsilon}{16}} - \sin^2 \theta_i}}, \\
       G_2  & = \int_{- 1}^{1} \varepsilon \sin \theta_i \sqrt{1 - e^{\frac{\varepsilon (8 + 9 z - 2 z^3)}{32}}\sin^2 \theta_i} \frac{\left(-21 z + 18 z^3\right)}{64} d z, \nonumber
\end{align}
as the particle crosses the interface and eventually to
\begin{align}
        \label{eqn:analytical leaving}
       \theta_{1f} & = \left. \theta_1 \right|_{z_c \to \infty} =  \frac{ G_3 + C }{\sqrt{1 - e^{\varepsilon}\sin^2 \theta_i}}, \\
       G_3  & = \int_{1}^{\infty} \varepsilon \sin \theta_i e^{ \frac{\varepsilon}{2}}\sqrt{1 - e^{\varepsilon(1 -\frac{1}{16 z^3})}\sin^2 \theta_i} \frac{\left(- 12 + 9 z^2\right)}{64 z^5} d z, \nonumber \\
       C & = \sqrt{\frac{e^{\frac{\varepsilon}{16}} - e^{\varepsilon}\sin^2 \theta_i}{e^{-\frac{15}{16}\varepsilon} - \sin^2 \theta_i}} (G_2 + G_1), \nonumber
\end{align}
as the particle departs away from the interface. Accounting for the leading order reorientation, the final orientation of pushers or pullers as they cross and go far ahead of the interface is
\begin{equation}
    \theta_f = \theta_{0f} + \beta \theta_{1f} + O\left(\beta^2, \varepsilon \right),
\end{equation}
where $\theta_{0f} = \left. \theta_0 \right|_{z_c \to \infty}$ follows from \eqref{eqn:SnellsFinal} as $\sin \theta_{0f} = e^{\varepsilon/2} \sin \theta_i$.

\bibliography{references}

\end{document}